 \documentclass[aps,prl,twocolumn,groupedaddress,amsmath,amssymb,showpacs]{revtex4}
\usepackage{graphicx}
\usepackage{amsmath}
\usepackage{dcolumn}
\usepackage{bm}

\begin{document}
\title{Role of spatial coherence in polarization tomography}
\author{A. Aiello}
\author{J.P. Woerdman}
\affiliation{Huygens Laboratory, Leiden University\\
P.O.\ Box 9504, 2300 RA Leiden, The Netherlands}
\begin{abstract}
We analyze an experimental setup in which a quasi-monochromatic
spatially coherent beam of light is used to probe a paraxial
optical scatterer. We discuss the effect of the spatial coherence
of the probe beam on the Mueller matrix representing the
scatterer. We show that according to the degree of spatial
coherence of the beam, the \emph{same} scattering system can be
represented by \emph{different} Mueller matrices. This result
should serve as a warning for experimentalists.
\end{abstract}

\maketitle


%
A great deal of literature exists about the phenomenological
treatment of polarization of light by matrix methods
\cite{KligerBook}. Among these methods, the Mueller-Stokes one is
perhaps the most utilized for the description of the interaction
between a quasi-monochromatic beam of light, hereafter denoted as
the probe, and a polarization-affecting medium, namely the
scattering system. In the Mueller-Stokes formalism the beam of
light is represented by four real numbers, namely the Stokes
parameters, and the scattering system is represented by a $4
\times 4$ real matrix, the Mueller matrix. Recently
\cite{Gopala98}, considerable efforts have been devoted to the
study of the connection between the algebraic properties of
Mueller matrices and the corresponding physical properties of the
systems they represent. However, in most contributions
\cite{KligerBook,Gopala98} emphasis is given to the mathematical
aspects of the problem rather than to the physical ones.

On the contrary, in the present Letter  we study how the spatial
coherence of the probe beam affects the physical properties of the
Mueller matrix representing a given scattering
 system. Using  Wolf's unified theory of coherence and
polarization of random electromagnetic beams \cite{Wolf03}, we
demonstrate that the \emph{same} scattering system can be
represented by  \emph{different} Mueller matrices, the difference
depending on the degree of spatial coherence of the probe beam.
Specifically, we show that an optical scatterer may behave either
as a non-depolarizing or a depolarizing system according to
whether the probe beam was completely spatially coherent or
completely spatially incoherent.
A typical experimental setup for polarization tomography consists
of five elementary units: The source $\mathcal{S}$, the
polarization-preparer $\mathcal{P}$, the scattering system
$\mathcal{M}$ \cite{Kim&Gil}, the polarization-analyzer
$\mathcal{A}$, and the detector $\mathcal{D}$ (see Fig. 1).
\begin{figure}[!hb]
\centerline{\includegraphics[width=8cm]{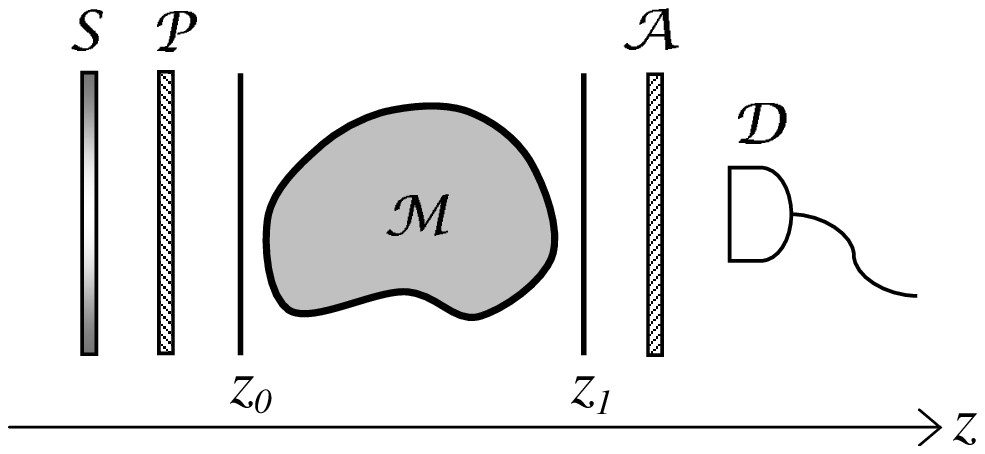}}
\caption{Polarization tomography setup described in the text. The
input $(z = z_0)$ and the output $(z = z_1)$ planes are indicated
by vertical lines.}
\end{figure}
The source $\mathcal{S}$ emits a quasi-monochromatic beam of light
at frequency $\omega$, which is prepared by $\mathcal{P}$ in a
well defined pure polarization state. This beam propagates close
to the $z$-axis through a scattering system characterized by its
spectral transmission matrix $G_{ij}(\mathbf{r}_1, \mathbf{r}_0,
\omega)$, $(i=x,y;\,j=x,y)$, where $\mathbf{r}_0=({\bm
\rho}_0,z_0)$, $\mathbf{r}_1=({\bm \rho}_1,z_1)$, and ${\bm
\rho}_A = (x_A, y_A), \,( A=0,1)$, are the transverse position
vectors on the input $(z=z_0)$ and output $(z = z_1)$ planes,
respectively.
Let
 $E_j(\mathbf{r}_0,\omega), \, (j=x,y)$ be a spectral Cartesian components of the
 electric field at the point $Q_0$ of coordinates
   $\mathbf{r}_0 = ({\bm \rho}_0,z_0), $ in the input  plane. The Cartesian components of the
 electric field at the point $P_1$ of coordinates $\mathbf{r}_1 = ({\bm \rho}_1, z_1)$ in the
 output  plane  are then given by:
\begin{equation}\label{eq10}
E_i(\mathbf{r}_1,\omega) = \int \mathrm{d }^2 \rho_0 \,
G_{ij}(\mathbf{r}_1, \mathbf{r}_0,
\omega)E_j(\mathbf{r}_0,\omega),
\end{equation}
 $(i=x,y;\,j=x,y)$, where  the integration extends over the transverse input-plane
coordinates: $\mathrm{d }^2 \rho_0 = \mathrm{d }x_0 \mathrm{d
}y_0$, and summation on repeated indices is understood. Behind the
scatterer, the polarization-analyzer $\mathcal{A}$ can select an
arbitrary field component $E(\mathbf{r}_1,\omega) = a_x
E_x(\mathbf{r}_1,\omega) + a_y E_y(\mathbf{r}_1,\omega)$, where
$a_x, a_y $ are complex amplitudes determined by the analyzer
setup.
 Finally, the detector $\mathcal{D}$ detects the spectral density
 $S(\mathbf{r}_1,\omega)$ (proportional to the averaged electric field density)
 of the beam at the point $P_1$:
$ S(\mathbf{r}_1,\omega) = \langle E(\mathbf{r}_1,\omega)
   E^*(\mathbf{r}_1,\omega) \rangle $, where angular brackets denote the average over an ensemble of
realizations of the scattered electric field  \cite{MandelBook}.
This expression for $S(\mathbf{r}_1,\omega)$ is a particular case
of the more general formula $S(\mathbf{r},\omega) = \mathrm{Tr}
\{\textsf{W}(\mathbf{r}, \omega)\}$ \cite{Mandel&Wolf}, where
$\mathrm{Tr}\{\textsf{W}(\mathbf{r}, \omega)\}$ denotes the trace
of the $2 \times 2$ spectral density matrix
$\textsf{W}(\mathbf{r}, \omega) \equiv \textsf{W}(\mathbf{r},
\mathbf{r},\omega)$ evaluated at the point $P$ (specified by the
position vector $\mathbf{r}$), and $\textsf{W}(\mathbf{r}_A,
\mathbf{r}_B,\omega)$ is the cross-spectral density matrix of the
beam at the pair of points $\mathbf{r}_A = ({\bm \rho}_A, z_0),
\mathbf{r}_B = ({\bm \rho}_A, z_0)$ in the plane $z = z_0$:
\begin{equation}\label{eq30}
  W_{ij}(\mathbf{r}_A,\mathbf{r}_B,\omega)  \equiv \langle E_i(\mathbf{r}_A,\omega)
   E_j^*(\mathbf{r}_B,\omega) \rangle,
\end{equation}
where $(i=x,y;j=x,y)$ \cite{Cross}.
We now consider the effects of the propagation through the
scattering system on the spectral density matrix
$\textsf{W}(\mathbf{r}, \omega)$ of the beam. From Eqs.
(\ref{eq10}) and (\ref{eq30}) it readily follows that
\begin{equation}\label{eq40}
\begin{array}{lcl}
\displaystyle{ \textsf{W}(\mathbf{r}_1,\omega) } & = &
\displaystyle{\int \mathrm{d }^2 \rho_0' \mathrm{d }^2 \rho_0''
\bigl\{ \textsf{G}(\mathbf{r}_1, \mathbf{r}_0', \omega) }
\\
&&   \displaystyle{ \times
\textsf{W}(\mathbf{r}_0',\mathbf{r}_0'',\omega)
 \textsf{G}(\mathbf{r}_0'', \mathbf{r}_1, \omega) \bigr\}},
 \end{array}
\end{equation}
where summation on repeated indices is understood and
$W_{kl}(\mathbf{r}_0',\mathbf{r}_0'',\omega) \equiv \langle
E_k(\mathbf{r}_0',\omega)
   E_l^* (\mathbf{r}_0'',\omega) \rangle$ are the elements of the cross-spectral density
   matrix $\textsf{W}(\mathbf{r}_0',\mathbf{r}_0'',\omega)$ of the beam at all pairs of points ${\bm \rho}_0',{\bm \rho}_0''$
   in the input plane, and  $\mathbf{r}_0' \equiv ({\bm \rho}_0',z_0)$, $\mathbf{r}_0''\equiv
({\bm \rho}_0'',z_0)$ respectively.
Furthermore, $ G_{lj}(\mathbf{r}_0'', \mathbf{r}_1, \omega)
  =  G_{jl}^*(\mathbf{r}_1,\mathbf{r}_0'',\omega)$. In order to obtain a
  formula involving Stokes parameters and Mueller matrices, we
  multiply both sides of Eq. (\ref{eq40}) by the normalized Pauli
  matrix
  $\sigma_{(\alpha)}$ $(\alpha = 0,1,2,3)$ \cite{AielloMath} and  trace, obtaining
\begin{equation}\label{eq50}
 \begin{array}{ll}
& \displaystyle{\mathcal{S}_\alpha(\mathbf{r}_1,\omega)  } \\ &
\displaystyle{ = \int \mathrm{d }^2 \rho_0' \mathrm{d }^2 \rho_0''
\mathcal{M}_{\alpha \beta} (\mathbf{r}_1,\mathbf{r}_0',
\mathbf{r}_0'', \omega) \mathcal{J}_\beta (\mathbf{r}_0',
\mathbf{r}_0'', \omega)},
 \end{array}
\end{equation}
$(\beta = 0,1,2,3)$, where summation on repeated indices is
understood and we have defined the spectral density Stokes
parameters \cite{Carozzi00} of the beam in the output plane as:
$\mathcal{S}_\alpha(\mathbf{r}_1,\omega) \equiv \mathrm{Tr}
\{\sigma_{(\alpha)} \textsf{W}(\mathbf{r}_1,\omega)\}$. Moreover,
we have introduced the \emph{cross-spectral density Mueller
matrix}
\begin{equation}\label{eq60}
 \begin{array}{ll}
& \displaystyle{\mathcal{M}_{\alpha \beta}
(\mathbf{r}_1,\mathbf{r}_0', \mathbf{r}_0'', \omega) } \\ &
\displaystyle{\equiv  \mathrm{Tr} \{\sigma_{(\alpha)}
\textsf{G}(\mathbf{r}_1,\mathbf{r}_0',\omega) \sigma_{(\beta)}
\textsf{G}^\dagger(\mathbf{r}_1,\mathbf{r}_0'',\omega)\}},
 \end{array}
\end{equation}
and the \emph{cross-spectral density Stokes parameters}  of the
input beam
\begin{equation}\label{eq70}
\mathcal{J}_\beta (\mathbf{r}_0', \mathbf{r}_0'', \omega) \equiv
\mathrm{Tr} \{\sigma_{(\beta)}
\textsf{W}(\mathbf{r}_0',\mathbf{r}_0'',\omega)\},
\end{equation}
which reduces to the input spectral density Stokes parameters
$\mathcal{S}_\beta(\mathbf{r}_0', \omega)$ in the limit
$\mathbf{r}_0'' \rightarrow \mathbf{r}_0'$.

Eq. (\ref{eq50})  shows that unlike the Stokes parameters
$\{S_\alpha\}$ in the elementary theory of partial polarization,
the spectral density Stokes parameters $\{\mathcal{S}_\alpha
(\mathbf{r}_1, \omega)\}$  at any point $P_1$ in the output plane
are not just a linear combination of the input spectral density
Stokes parameters $\{\mathcal{S}_\alpha (\mathbf{r}_0, \omega)\}$
but they are expressed in terms of the cross-spectral density
Stokes parameters $\{\mathcal{J}_\beta (\mathbf{r}_0',
\mathbf{r}_0'', \omega)\}$ at \emph{all} pairs of points ${\bm
\rho}_0',{\bm \rho}_0''$ in the input plane. This fact obviously
represents the effect of the coherence of the input beam.
Moreover, since for $\mathbf{r}_0' \neq \mathbf{r}_0''$,
$\textsf{G}^\dagger(\mathbf{r}_1, \mathbf{r}_0'', \omega) \neq
[\textsf{G}( \mathbf{r}_1, \mathbf{r}_0', \omega)]^\dagger$, the
cross-spectral density Mueller matrix is \emph{not} a
Mueller-Jones matrix \cite{Gopala98}.
So far we have been concerned with a generic spatially coherent
light source, without specifying its degree of coherence. We shall
now consider with some detail the two opposite limit cases of
completely coherent and completely incoherent light. For both
cases we consider a uniformly polarized input beam  specified by
the coordinate- and frequency-independent two-dimensional unit
vector ${\bm e} = (e_x, e_y)$, such that $
E_i(\mathbf{r}_0,\omega) = e_i E(\mathbf{r}_0, \omega),
\,(i=x,y)$, where $E(\mathbf{r}_0, \omega)$ is a scalar function
of the point $Q_0$ in the input plane. Using this assumption, one
readily obtains the following expression  for the cross-spectral
density matrix of the input beam:
\begin{equation}\label{eq80}
W_{kl} (\mathbf{r}_0',\mathbf{r}_0'',\omega) = \mathcal{E}_{kl}
w(\mathbf{r}_0',\mathbf{r}_0'',\omega),
\end{equation}
where $\mathcal{E}_{kl} \equiv e_k e_l^*$, and
$w(\mathbf{r}_0',\mathbf{r}_0'',\omega) \equiv \left\langle
E(\mathbf{r}_0', \omega) E^*(\mathbf{r}_0'', \omega)
\right\rangle$ is the scalar cross-spectral density function which
characterize the second-order coherence properties of the input
beam. If we substitute  Eq. (\ref{eq80}) into Eq. (\ref{eq50}) and
use  Eqs. (\ref{eq60}-\ref{eq70}) we obtain
\begin{equation}\label{eq90}
\begin{array}{lcl}
\displaystyle{ S_{\alpha}(\mathbf{r}_1,\omega)} & = &
\displaystyle{\int \mathrm{d }^2 \rho_0' \mathrm{d }^2 \rho_0''
w(\mathbf{r}_0',\mathbf{r}_0'',\omega)  }
\\&&
\displaystyle{ \times \mathcal{M}_{\alpha \beta}
(\mathbf{r}_1,\mathbf{r}_0', \mathbf{r}_0'', \omega) S_{\beta}  },
 \end{array}
\end{equation}
where $ S_{\beta}  \equiv \mathrm{Tr} \{\sigma_{(\beta)}
\mathcal{E} \}$, $(\beta = 0,1,2,3)$, are the input Stokes
parameters which are independent from both the input-plane
coordinates $\mathbf{r}_0$ and the frequency $\omega$.
In the case of a completely coherent source
\begin{equation}\label{eq100}
w(\mathbf{r}_0',\mathbf{r}_0'',\omega) = u(\mathbf{r}_0',\omega)
u^*(\mathbf{r}_0'',\omega),
\end{equation}
where $u(\mathbf{r}_0,\omega)$ represents the complex disturbance
of the field in the input plane \cite{BornWolf}. From Eqs.
(\ref{eq90}) and (\ref{eq100}) readily follows that
$S_{\alpha}(\mathbf{r}_1,\omega) = M^{(C)}_{\alpha \beta}
(\mathbf{r}_1,\omega) S_{\beta} $ where the ``coherent'' Mueller
matrix $\textsf{M}^{(C)} (\mathbf{r}_1,\omega)$ at the output
point $P_1$ is defined as
\begin{equation}\label{eq110}
 M^{(C)}_{\alpha \beta}
(\mathbf{r}_1,\omega) \equiv \mathrm{Tr} \{\sigma_{(\alpha)}
\textsf{K}(\mathbf{r}_1, \omega) \sigma_{(\beta)}
\textsf{K}^\dagger(\mathbf{r}_1, \omega) \},
\end{equation}
where $ \textsf{K}(\mathbf{r}_1, \omega) \equiv \int \mathrm{d }^2
\rho_0 \, u(\mathbf{r}_0,\omega)\textsf{G}(\mathbf{r}_1,
\mathbf{r}_0, \omega)$ represents the tensor-valued complex
disturbance of the field in the output plane. Eq. (\ref{eq110})
shows that in the ideal case of a completely coherent and
uniformly polarized probe beam, the  Mueller matrix $
M^{(C)}_{\alpha \beta} (\mathbf{r}_1,\omega)$ is a Mueller-Jones
matrix \cite{Gopala98} representing a \emph{non-depolarizing}
scattering system.
In the case of a completely incoherent source
\begin{equation}\label{eq120}
w(\mathbf{r}_0',\mathbf{r}_0'',\omega) = w(\mathbf{r}_0',\omega)
\delta^{(2)}({\bm \rho}_0' - {\bm \rho}_0'' ),
\end{equation}
where $w(\mathbf{r}_0',\omega) \geq 0$ is the spectral density
function of the input beam. If we substitute Eq. (\ref{eq120}) in
Eq. (\ref{eq90}) we obtain $S_{\alpha}(\mathbf{r}_1,\omega) =
M^{(I)}_{\alpha \beta} (\mathbf{r}_1,\omega) S_{\beta} $ where the
``incoherent'' Mueller matrix $\textsf{M}^{(I)}
(\mathbf{r}_1,\omega)$ at the output point $P_1$ is defined as
\begin{equation}\label{eq130}
 M^{(I)}_{\alpha \beta} (\mathbf{r}_1,\omega)   =  \int
\mathrm{d }^2 \rho_0' w(\mathbf{r}_0',\omega) \mathcal{M}_{\alpha
\beta} (\mathbf{r}_1,\mathbf{r}_0', \omega) ,
\end{equation}
where the spectral density Mueller matrix $\mathcal{M}_{\alpha
\beta} (\mathbf{r}_1,\mathbf{r}_0', \omega) \equiv
\mathcal{M}_{\alpha \beta} (\mathbf{r}_1,\mathbf{r}_0',
\mathbf{r}_0', \omega)$ is now a bona fide Mueller-Jones matrix.
Eq. (\ref{eq130}) expresses the incoherent Mueller matrix
$\textsf{M}^{(I)} (\mathbf{r}_1,\omega)$ as a linear combination,
with non-negative coefficients $w(\mathbf{r}_0',\omega)$, of the
Mueller-Jones matrices $\mathcal{M}_{\alpha \beta}
(\mathbf{r}_1,\mathbf{r}_0', \omega)$. Therefore $\textsf{M}^{(I)}
(\mathbf{r}_1,\omega)$ is formally equivalent to a Mueller matrix
representing a \emph{depolarizing} scattering system
\cite{Kim&Gil}. Moreover, from Eq. (\ref{eq130}) readily follows
that the Hermitian matrix $\textsf{H}$ \cite{Anderson94}
associated with $\textsf{M}^{(I)} (\mathbf{r}_1,\omega)$ is
positive semidefinite, namely $\textsf{M}^{(I)}
(\mathbf{r}_1,\omega)$ satisfy the so called Jones criterion and
can be referred to as a ``physical'' Mueller matrix
\cite{Gopala98}.
Eqs. (\ref{eq110}) and (\ref{eq130}) are the key results of this
Letter: they give the Mueller matrices
$\textsf{M}^{(C)}(\mathbf{r}_1,\omega)$ and
$\textsf{M}^{(I)}(\mathbf{r}_1,\omega)$ describing the \emph{same}
scattering system probed by either a completely coherent or a
completely incoherent light beam, respectively. Surprisingly, we
have found that in the first case
$\textsf{M}^{(C)}(\mathbf{r}_1,\omega)$ is a Mueller-Jones matrix
which represents a non-depolarizing optical system,  while in the
second case $\textsf{M}^{(I)}(\mathbf{r}_1,\omega)$ is a
Jones-derived (or ``physical'') Mueller matrix that represents a
depolarizing optical system.

The more general (and realistic) case of partially coherent
quasi-monochromatic light probe is illustrated by Eqs.
(\ref{eq50}-\ref{eq70},\ref{eq90}). In particular Eq. (\ref{eq60})
shows that when a partially coherent probe beam is used, the
cross-spectral density Mueller matrix may not be interpreted as a
Mueller-Jones matrix. However, since in general the cross-spectral
density function can be expanded as a superposition of coherent
fields $\{u_n({\bm \rho}_0,\omega)\}$  as
$w(\mathbf{r}_0',\mathbf{r}_0'',\omega) = \sum_{n} \gamma_n (z_0,
\omega) u_n({\bm \rho}_0',\omega)u_n^*({\bm \rho}_0'',\omega)$
\cite{MandelBook}, where $\gamma_n (z_0, \omega) \geq 0$,
 it is easy to see that the ``partially-coherent'' Mueller
matrix $\textsf{M}^{(P)}(\mathbf{r}_1,\omega)$
\begin{equation}\label{eq140}
\begin{array}{lcl}
\displaystyle{\mathcal{M}^{(P)}_{\alpha
\beta}(\mathbf{r}_1,\omega)} & = & \displaystyle{\int \mathrm{d
}^2 \rho_0' \mathrm{d }^2 \rho_0'' \,
w(\mathbf{r}_0',\mathbf{r}_0'',\omega)  }
\\&&
\displaystyle{ \times \mathcal{M}_{\alpha \beta}
(\mathbf{r}_1,\mathbf{r}_0', \mathbf{r}_0'', \omega) },
 \end{array}
\end{equation}
can be also decomposed as a superposition [with non-negative
coefficients $\gamma_n (z_0, \omega)$] of ``coherent''
Mueller-Jones matrices $\textsf{M}^{(n)}(\mathbf{r}_1,\omega) =
\mathrm{Tr} \{\sigma_{(\alpha)} \textsf{K}_n(\mathbf{r}_1, \omega)
\sigma_{(\beta)} \textsf{K}^\dagger_n(\mathbf{r}_1, \omega) \}$
where $ \textsf{K}_n(\mathbf{r}_1, \omega) \equiv \int \mathrm{d
}^2 \rho_0 \, u_n(\mathbf{r}_0,\omega)\textsf{G}(\mathbf{r}_1,
\mathbf{r}_0, \omega)$. Therefore the matrix
$\textsf{M}^{(P)}(\mathbf{r}_1,\omega)$ represents, in general, a
depolarizing optical system. It is worthwhile to note that each
matrix $\textsf{M}^{(n)}(\mathbf{r}_1,\omega)$ depends in a
nontrivial way on the field-amplitudes $ u_n(\mathbf{r}_0,\omega)$
of the probe beam. Therefore, when using spatially coherent probe
beams, the \emph{linearity} required by the Mueller-Stokes
formalism can be ensured only with respect to the polarization
degrees of freedom of the beam (viz, the matrix indices), but not
with respect to the spatial degrees of freedom.

In conclusion, we have shown that, contrary to common belief, the
Mueller matrix representing an optical system in a scattering
experiment, is not determined by the system only but strongly
depends on the spatial coherence of the probe beam. This fact
poses serious limitations to the range of applicability of the
Mueller-Stokes formalism. Consider, for instance, the following
case: a probe beam propagates through a medium suffering multiple
scattering. Since the degree of spatial coherence of the  beam
changes during  propagation \cite{Wolf03b}, the last part of the
medium will be probed by a beam qualitatively different from the
one which probed the first part of the medium. Therefore the
Mueller matrix representing the medium will not only be determined
by the polarization properties of the medium itself, but also by
the way the medium affects the coherence properties of the probe
beam. With respect to this problem, the use of completely
spatially incoherent probe beams for polarization tomography seems
highly preferable.

 Dirk Voigt, Graciana Puentes and Martin van Exter are acknowledged for
 valuable discussions. We also acknowledge support from the EU under the
IST-ATESIT contract. This project is also supported by FOM.


\begin{thebibliography}{abcd22}
\bibitem{Cross}
For the physical meaning of $\textsf{W}(\mathbf{r}_A,
\mathbf{r}_B,\omega)$ and its relation with the $2 \times 2$
coherence matrix $J$ used in elementary theory of partial
polarization, see Ref. \cite{Wolf03}, pag. 264.
\end{thebibliography}
\end{document}